\documentclass[aps,a4paper,twocolumn]{revtex4-2}
\usepackage{amsmath}
\usepackage{graphicx}
\usepackage{subfigure}
\usepackage[usenames,dvipsnames]{color}
\definecolor{darkblue}{RGB}{0,0,196}
\usepackage[colorlinks=true,linkcolor=darkblue,citecolor=darkblue,urlcolor=darkblue]{hyperref}
\usepackage{setspace}
\usepackage{hyperref}
\usepackage{xcolor}
\hypersetup{
  colorlinks   = true, 
  urlcolor     = red, 
  linkcolor    = blue, 
  citecolor   = blue 
}
\usepackage{graphicx}
\usepackage{amsmath,bbm}
\usepackage{amssymb,bm}
\usepackage{yfonts}
\usepackage{comment}
\usepackage{tabularx}

\usepackage{booktabs} 
\usepackage{float}    
\usepackage{placeins} 

\usepackage[usenames,dvipsnames]{color}
\definecolor{darkblue}{RGB}{0,0,196}
\usepackage[colorlinks=true,linkcolor=darkblue,citecolor=darkblue,urlcolor=darkblue]{hyperref}

\begin{document}
\title{Finite $V_{\rm 2\Delta}$ puzzle in low-multiplicity pp collisions from ultra-long-range azimuthal correlations in the string-shoving model}

\author{Antonio Ortiz}%
\author{Dushmanta Sahu}
 \email{Dushmanta.Sahu@cern.ch}
\affiliation{%
 Instituto de Ciencias Nucleares, UNAM\\
 Circuito exterior s/n Ciudad Universitaria, 04510 Mexico City, Mexico
}%
\author{Gyula Bencedi} 

\affiliation{HUN-REN Wigner Research Centre for Physics, P.O. Box 49, H-1525 Budapest}

\begin{abstract}

Ultra-long range angular correlations have been recently reported by the ALICE collaboration in pp collisions at $\sqrt{s}=13$ TeV below ${\rm d}N_{\rm ch}/{\rm d}\eta=7$. The measurements have been performed as a function of the charged-particle multiplicity at midrapidity ($N_{\rm ch}$ in $|\eta|<0.8$), which is known to be strongly sensitive to local multiplicity fluctuations. The present work investigates the impact of the event-activity estimator on ultra-long range angular correlations. The study is conducted in the framework of PYTHIA8 with the string shoving mechanism since it gives a non-zero elliptic flow coefficient, $V_{2\Delta}$. Such collective-like effects arise from repulsive interactions between overlapping color strings, without invoking hydrodynamic collectivity. The analysis is conducted as a function of $N_{\rm ch}$, the number of parton-parton scatterings ($N_{\rm mpi}$) and flattenicity. Surprisingly, for ultra-long range correlations, pp collisions with $N_{\rm mpi}=1$ (dijets) seems to be the most sensitive to string shoving. The effect diminishes with increasing $N_{\rm mpi}$. While in data, within uncertainties, $V_{2\Delta}$ exhibits a weak multiplicity dependence; the string shoving mechanism gives a $V_{2\Delta}$ that decreases with the increase in $N_{\rm ch}$. The present work therefore supports the picture stating that mechanisms such as string shoving might explain the low multiplicity limit, whereas, hydro becomes relevant in high-multiplicity pp collisions. This work also suggests that flattenicity might be more effective than $N_{\rm ch}$ to better handle non-flow effects.
 
\end{abstract}
\maketitle
\section{Introduction}

The study of the quark--gluon plasma (QGP), which is a deconfined state of quarks and gluons that exists at extremely high temperatures and energy densities, is one of the most exciting areas of high-energy physics research currently. In heavy-ion collisions at the Relativistic Heavy Ion Collider (RHIC) and the Large Hadron Collider (LHC), compelling evidence for QGP formation have emerged through multiple observables, including strong jet quenching~\cite{CMS:2016xef,ALICE:2018vuu}, strangeness enhancement~\cite{ALICE:2013xmt}, charmonia suppression~\cite{ALICE:2023gco} and particularly, the observation of collective anisotropic flow~\cite{CMS:2011cqy,ALICE:2011svq}. Anisotropic flow is observed as momentum-space anisotropies in the final-state particle distributions~\cite{Ollitrault:1992bk,Voloshin:1994mz}. It arises from the conversion of the initial spatial asymmetries of the overlapping region in colliding nuclei into momentum anisotropies via collective interactions in the hot, dense medium. The second-order Fourier coefficient $v_2$, known as elliptic flow, has proven to be one of the most sensitive probes of the transport properties of the QGP \cite{STAR:2000ekf}. In non-central heavy-ion collisions, the initial spatial anisotropy of the overlap region gets transformed into momentum-space anisotropy via multiple rescatterings, with the magnitude of $v_2$ being particularly sensitive to the shear viscosity to entropy density ratio $\eta/s$ of the created medium \cite{Romatschke:2007mq,Song:2007ux,Hirano:2005wx}.

The hydrodynamic model, which successfully describes the large collective effects observed in heavy-ion collisions, was traditionally expected to be inapplicable to small systems such as proton-proton (pp) and proton-nucleus (p--Pb) collisions due to their limited system size and short lifetime. Traditionally, the view was that only high energy heavy-ion collisions could produce the thermalized fluid, and small systems were only useful as the baseline for the QGP studies. However, the unexpected observation of long-range ridge-like correlations in high-multiplicity pp~\cite{CMS:2016fnw} and p--Pb~\cite{CMS:2016wfo} collisions at the LHC challenged this conventional view. These observations revealed collective-like behavior in high-multiplicity pp collisions, strikingly similar to that observed in heavy-ion collisions, including the mass-ordering of $v_2$ for identified particles. In addition to this, other QGP-like signatures such as strangeness enhancement \cite{ALICE:2016fzo,ALICE:2020nkc} have also been experimentally observed in high-multiplicity pp collisions, which makes the small system QGP an interesting domain of study. Moreover, high-multiplicity pp collisions show thermodynamic and transport properties equivalent to peripheral heavy-ion collisions~\cite{Sahu:2020nbu,Sahu:2020mzo,Mishra:2021yer,Sahu:2020swd}. Some studies have shown that hydrodynamics can also be applied to such collisions after certain charged particle multiplicity (${\rm d}N_{\rm ch}/{\rm d}\eta \simeq 10$) ~\cite{Scaria:2022yrz,Habich:2015rtj,Heinz:2019dbd,Hatwar:2022cbx}, suggesting a strong possibility of thermalized medium production in small systems. However, QGP-like signals have been recently reported in pp collisions at lower multiplicities~\cite{ALICE:2025bwp}, and in  ultra-small systems such as jets~\cite{CMS:2023iam,Vertesi:2024fwl} and ultra-peripheral collisions.

The discovery of collective phenomena in small systems has sparked intense debate regarding their origin. Two principal explanations have emerged; final-state effects involving the formation of a small droplet of QGP with hydrodynamic behavior, and initial-state effects arising from the Color Glass Condensate (CGC) effective theory describing the high-energy wavefunctions of the colliding hadrons \cite{Dusling:2015rja}. Recent ALICE measurements~\cite{ALICE:2025bwp} of ultra-long-range correlations with pseudorapidity separations of $|\Delta\eta| > 5.0$ in pp and $|\Delta\eta| > 6.5$ in p--Pb collisions down to minimum-bias multiplicities have revealed significant discrepancies with both hydrodynamic and CGC-based calculations \cite{Zhao:2022ugy}. The 3D-Glauber + MUSIC + UrQMD hydrodynamic model severely underestimates the calculated $V_{2\Delta}$ values at low multiplicities, while CGC calculations generate only short-range correlations that vanish at ultra-long ranges~\cite{Schenke:2022mjv}. 

In the midst of theoretical challenges, alternative mechanisms implemented in Monte Carlo event generators offer compelling explanations. The PYTHIA8 event generator \cite{Sjostrand:2007gs,Sjostrand:2014zea}, which successfully describes a wide range of pp collision observables, incorporates the string shoving mechanism \cite{Bierlich:2017vhg,Bierlich:2016smv} as a potential explanation for collective-like effects without invoking QGP formation. In this model, two or more overlapping color strings generated by partonic interactions exert transverse pressure on each other, creating a collective push that mimics hydrodynamic behavior~\cite{Bierlich:2024odg}. However, recent comparisons with ALICE data \cite{ALICE:2025bwp} have revealed significant limitations of the string shoving approach. While it can generate ridge-like structures in high-multiplicity events, it produces a decreasing trend of $v_2$ with increasing $N_{\text{ch}}$, opposite to the experimentally observed behavior. 

In view of such limited understanding, this work presents a systematic investigation of the string shoving mechanism in PYTHIA8, focusing on the origin of finite $v_2$ in the low multiplicity limit. In this context, the model is used as a proxy to study collective-like effects in ultra-small systems. The conclusions might be relevant for the study of collectivity using the LHC data. The dependence of two-particle correlations is examined on $N_{\text{ch}}$, number of parton-parton scatterings ($N_{\text{mpi}}$), and event shape characterized by flattenicity~\cite{Ortiz:2022mfv}. Crucial tests of the string shoving mechanism are provided and its limitations in describing collective phenomena in small-collision systems are discussed. Moreover, comments are made on how charged-particle multiplicity introduces biases that might overestimate the two-particle correlation and how flattenicity does a better job.

The paper is organized as follows. Section~\ref{methodology} describes the analysis framework, including event generation, observable definitions, and the correlation analysis technique. Section~\ref{results} presents the main findings on the $N_{\text{ch}}$, $N_{\text{mpi}}$, and flattenicity dependence of correlation study and discusses the implications of the results, comparing them with experimental data and theoretical expectations. Finally, section~\ref{summary} summarizes the main conclusions and outlines future research directions.

\section{Methodology}
\label{methodology}

\subsection{PYTHIA8 Event Generator}
PYTHIA8 is a Monte Carlo event generator for high-energy particle collisions that implements a comprehensive framework for simulating proton-proton, proton-antiproton, and electron-positron collisions~\cite{Sjostrand:2007gs,Sjostrand:2014zea}. PYTHIA Angantyr is a heavy-ion extension to PYTHIA8, which models proton-nucleus (p–A) and nucleus-nucleus (A–A) collisions by extrapolating and generalizing the physics models from pp collisions~\cite{Bierlich:2018xfw}. The generator incorporates multiple physics mechanisms including parton distribution functions, initial-state and final-state parton showers, multiparton interactions (MPI), and hadronization processes. The core of PYTHIA8's approach to hadronization is based on the Lund string model, where colored partons are connected by color flux tubes that fragment into hadrons according to the Schwinger mechanism~\cite{Andersson:1983ia,Sjostrand:1982fn}. The event generation in PYTHIA8 proceeds through several sequential stages; hard process generation via matrix elements, initial-state radiation, multiparton interactions, final-state radiation, beam remnants handling, and finally hadronization. For minimum-bias pp collisions and underlying event studies, PYTHIA8 includes models for non-perturbative quantum chromodynamic (QCD) processes through its multiparton interactions framework, which accounts for the fact that high-energy collisions often involve several parton-parton scatterings in addition to the primary hard processes. The generator has been extensively tuned to reproduce a wide range of experimental data from LEP, Tevatron, and LHC experiments, making it a standard tool for simulating jet-dominated processes in proton-proton collisions. For this analysis, PYTHIA8 (version 8.312) is used for all the simulations.

PYTHIA8 Monash tune serves as a standard benchmark for simulating proton-proton collisions. This tune provides a comprehensive calibration of the underlying physics models to reproduce a broad set of experimental data across various collision energies~\cite{Skands:2014pea}. Furthermore, the tune modifies parameters of the Lund string fragmentation model, which affects baryon production rates. However, despite these adjustments, it still struggles to quantitatively reproduce the multiplicity-dependent enhancement of strange and multi-strange hadron production observed in LHC data. Nonetheless, the use of the Monash tune ensures a reliable and well-validated baseline description of the collision environment, against which the specific effects of other mechanisms can be cleanly isolated and studied.

\begin{table*}[t!]
\centering
\caption{List of the relevant parameters of the PYTHIA 8 settings used in the current work.}
{\renewcommand{\arraystretch}{1.2}
\begin{tabular}{|l@{\hspace{2cm}}|c@{\hspace{2cm}}|}
\hline
\textbf{PYTHIA8 Parameter} & \textbf{Settings} \\
\hline
\hline
\texttt{Beams:eCM} & 13000 \\
\texttt{SoftQCD:nonDiffractive} & on \\
\hline
\texttt{Ropewalk:doShoving} & on \\
\texttt{Ropewalk:RopeHadronization} & on \\
\texttt{Ropewalk:gAmplitude} & 10\\
\texttt{Ropewalk:r0} & 0.41 \\
\texttt{Ropewalk:rCutOff} & 10\\
\texttt{Ropewalk:deltay} & 0.10 \\
\hline
\texttt{ParticleDecays:limitTau0} & on \\
\texttt{PartonVertex:setVertex} & on \\
\texttt{PartonVertex:ProtonRadius} & 0.70 \\
\hline
\end{tabular}}
\label{tab1}
\end{table*}

The string shoving mechanism is an extension to the standard PYTHIA8 framework that aims to explain collective-like effects in high-multiplicity proton-proton collisions without invoking QGP formation~\cite{Bierlich:2016vgw}. In this approach, multiparton interactions create a dense system of overlapping color strings during the initial stages of the collision. When these strings overlap in the transverse plane, they generate a repulsive interaction pressure due to the exchange of virtual gluons between adjacent strings. The strength of this shoving effect is governed by a dimensionless parameter (Ropewalk:gAmplitude) $g$ that controls the repulsive interaction strength between overlapping strings. In the recent ALICE study, two extreme values are used; $g = 3$ representing moderate shoving strength and $g = 40$ corresponding to strong repulsive interactions~\cite{ALICE:2025bwp}. The transverse pressure generated by string shoving mimics some aspects of transverse flow observed in heavy-ion collisions, particularly the development of ridge structures in two-particle correlations. The pseudorapidity dependence of the shoving effect is directly linked to the spatial extension of the color strings in the longitudinal direction. In the string shoving framework, the observed long-range correlations arise from the initial geometry of the overlapping string system rather than from hydrodynamic evolution of a thermalized medium. 

For this analysis, 10 billion pp collisions are generated with the string shoving mechanism. For a conservative estimation, the string shoving parameter value $g$ = 10 is chosen. With the generated events, the study proceeds to use the template fit method and extract the final second order two-particle correlation. Table~\ref{tab1} shows the relevant parameters which we have used for the simulations. The rest of the parameters are set as default.

\subsection{Template Fit Method for Non-Flow Subtraction}

For this analysis, the required per-trigger yield contains contributions from both genuine flow correlations and non-flow effects arising from jets and resonance decays. To isolate the flow component and minimize non-flow contamination, the template-fit method is applied to the $\Delta\varphi$ projections of the two-dimensional yield. However, instead of applying per-trigger normalization, we rescale each $\Delta\varphi$ distribution by its integral, resulting in distributions with unit area. This normalization preserves the relative modulation pattern while removing trivial variations in absolute yield so that we can observe the differences in shapes clearly. Thus, the $\Delta\varphi$ projections are computed as~\cite{ATLAS:2015hzw},

\begin{equation}
Y(\Delta\varphi) = \frac{\mathrm{d}N_{\rm pair}}{\mathrm{d}\Delta\varphi} 
= 
\int \left(\frac{\mathrm{d}^{2}N_{\rm pair}}{\mathrm{d}\Delta\eta\mathrm{d}\Delta\varphi}\right) \mathrm{d}\Delta\eta.
\end{equation}

Here, $\frac{dN_{\text{pair}}}{d\Delta\varphi}$ is the integrated pair distribution. For the template fit, we further normalize this distribution to unit area:
\begin{equation}
Y(\Delta\varphi) = \frac{\frac{dN_{\text{pair}}}{d\Delta\varphi}}{\int \frac{dN_{\text{pair}}}{d\Delta\varphi} \, d\Delta\varphi},
\end{equation}
where $Y(\Delta\varphi)$ is the observable used in Eq.~(2).  The template-fit procedure assumes that higher multiplicity events can be expressed as a superposition of low-multiplicity events with additional flow contributions. Thus the fitting function takes the form~\cite{ALICE:2013snk,CMS:2012qk},

\begin{equation}
Y(\Delta\varphi) = F Y^{\rm LM}(\Delta\varphi) + G  \left[1 + \sum_{n=2}^{3} 2V_{n\Delta} \cos(n\Delta\varphi)\right],
\end{equation}

where $Y(\Delta\varphi)$ and $Y^{\rm LM}(\Delta\varphi)$ represent the one-dimensional $\Delta\varphi$ projections in the high-multiplicity and low-multiplicity event classes, respectively. The parameters $F$ and $G$ are scaling factors, while the Fourier series describes the additional flow contribution in high-multiplicity event classes. The two-particle correlation coefficients $V_{n\Delta}$ are extracted by fitting the $\Delta\varphi$ projections of high-multiplicity event classes using this equation. Here, $n$ = 2 gives the second-order coefficient of the two-particle correlation. We note that the extracted $V_{2\Delta}$ is independent of the choice of normalization. If one instead uses the conventional per-trigger yield $\frac{1}{N_{\text{trig}}} \frac{dN_{\text{pair}}}{d\Delta\varphi}$, the only difference is an overall constant factor relative to $Y(\Delta\varphi)$. Because the template fit in Eq.~(2) contains free scaling parameters $F$ and $G$, this constant factor is absorbed, leaving $V_{2\Delta}$ unchanged. Thus, our results are directly comparable to experimental measurements that use per-trigger normalization.

The two-particle correlation function is constructed using charged particles within the forward ($2.6 < \eta < 3.2$) and backward ($-3.0 < \eta < -2.0$) pseudorapidity intervals. Particle pairs are formed by combining one particle from the forward interval and one from the backward interval. For each candidate pair, $\Delta\eta = \eta_{1} - \eta_{2}$ is calculated, and only pairs satisfying $5.0 < |\Delta\eta| < 6.0$ are retained for the correlation analysis. In Monte Carlo studies with perfect detector acceptance, the primary acceptance effect is the limited $\eta$ range, which is already enforced by the pair selection above. To verify that no further acceptance corrections are needed, we employ a randomized background method as a cross-check. For each event, we construct a randomized background by independently assigning random $\varphi$ and $\eta$ values to all particles, destroying physical correlations. The resulting randomized background $B(\Delta\varphi)$ is found to be flat within statistical fluctuations, confirming that the raw signal distribution $S(\Delta\varphi)$ already contains the full physical correlation information. Since an overall scale factor does not affect the extracted Fourier coefficients $V_{n\Delta}$, we perform the template fit directly on $S(\Delta\varphi)$. This approach is equivalent to the $S/B$ ratio method generally used in experimental analyses, but eliminates unnecessary computational steps.

The ultra-long-range correlations with $|\Delta\eta| > 5.0$ in pp collisions are supposed to be not affected by non-flow contributions from relatively short-range resonance decays and jet fragmentation on the near side ($\Delta\varphi < \pi/2$) of the yield. Jet contributions on the away side ($\Delta\varphi > \pi/2$) are accounted for using the non-flow subtraction with the template fit method to extract $V_{2\Delta}$. The template-fit method assumes that the non-flow component of the correlation function retains a similar shape across multiplicity classes. This assumption is particularly well justified at \(|\Delta\eta| > 5\) \cite{ALICE:2023lyr, ATLAS:2019xqc}, where jet-induced correlations are strongly suppressed due to the large rapidity gap, resonance decays, Bose--Einstein and Coulomb effects are negligible over such large pseudorapidity separations. Consequently, any residual non-flow at ultra-long range is expected to be small and slowly varying in \(\Delta\varphi\), making the template method especially suitable for suppressing short-range non-flow contributions and extracting the dominant long-range azimuthal modulations.

The current $V_{2\Delta}$ parameter differs from the $v_2$ observable generally reported in literature, which is estimated using the three pairs of two-particle correlations method. This procedure partially cancels out the effects of longitudinal flow decorrelation. However, this cancellation effect makes $v_2$ less sensitive to flow decorrelations, potentially leading to biased results especially at long-range. In contrast, the $V_{2\Delta}$ observable used in the current analysis is more sensitive to longitudinal flow decorrelation effects and does not rely on any assumptions related to flow factorization, thereby providing tighter constraints on theoretical models. This enhanced sensitivity makes it particularly suitable for investigating ultra-long-range correlations and their dependence on collision multiplicity and pseudorapidity separation. The extraction of $V_{2\Delta}$ through the template fit method is assumed to provide a direct quantification of the two-particle correlation strength that is minimally affected by non-flow effects and offers strong resolving power between different physical mechanisms contributing to long-range correlations in small collision systems.

To validate our implementation of the template-fit procedure, we reproduce the PYTHIA8 results reported by the ALICE collaboration in Ref.~\cite{ALICE:2025bwp} using the identical generator settings and kinematic cuts. Our extracted $V_{2\Delta}$ values for the multiplicity classes agree with the ALICE-reported PYTHIA8 results well within uncertainties. This cross-check confirms that our implementation, including the low-multiplicity template definition, fitting algorithm, and uncertainty propagation, is correct and unbiased.

Throughout this paper, 'non-flow' refers conventionally to correlations from jets, resonance decays, and quantum statistics. The template-fit method
subtracts such contributions. PYTHIA8 with string shoving generates azimuthal correlations without a thermalized medium. This comes under 'collective-like' effect, which denotes any azimuthal modulation resembling flow without implying hydrodynamics. This distinction allows model comparisons without theoretical prejudice.

\subsection{Event activity estimators}

The charged-particle multiplicity at mid-rapidity, $N_{\rm ch}$, serves as the most traditional and widely used event-activity estimator in small-collision systems. It is defined as the number of charged particles estimated within a narrow pseudorapidity interval. While experimentally easy to estimate, $N_{\rm ch}$ possesses a significant limitation for studies of long-range collective phenomena. It is highly sensitive to local, hard processes such as jet production and resonance decays (see e.g.~\cite{Mendez:2025dqz}). An event with a single high-transverse-momentum jet can yield a high $N_{\rm ch}$ value without a correspondingly high density of underlying soft processes or MPIs. This introduces a strong auto-correlation bias when the observable of interest, such as $V_{2\Delta}$ is estimated. Consequently, event classes defined by $N_{\rm ch}$ can conflate topologically distinct event types, mixing isotropic, MPI-rich events with jet-quenched, pencil-like events. This potentially obscures or distorts the genuine correlation between the initial-state geometry and the final-state azimuthal anisotropy.

In the PYTHIA8 event generator, MPI represent a foundational mechanism wherein multiple independent parton-parton scatterings occur within a single proton-proton collision, contributing significantly to the underlying event. These interactions are predominantly semi-hard, involving momentum transfers above a few GeV/$c$, and are crucial for describing the observed charged-particle multiplicity and transverse momentum spectra in minimum-bias collisions. The MPI framework is integrated with models for initial and final-state radiation and hadronization via the Lund string fragmentation. Furthermore, the phenomenon of color reconnection (CR), where the color charges of final-state partons are rearranged to reduce the total string length, is intrinsically linked to MPI. In high-multiplicity events, a large number of MPI leads to overlapping strings, which can form color ropes in advanced hadronization models, enhancing the production of strange hadrons and mimicking collective effects such as radial flow, thereby providing a non-plasma explanation for fluid-like behavior observed in small-collision systems~\cite{OrtizVelasquez:2013ofg}.

Flattenicity is a novel event classifier introduced to characterize the global topology of proton-proton collisions, with a specific design goal to isolate events dominated by MPI~\cite{Ortiz:2022mfv,ALICE:2024vaf}. The experimentally accessible definition of flattenicity $\rho_{\mathrm{nch}}$ is given as~\cite{Ortiz:2022mfv}; 

\begin{equation}
\rho_{\mathrm{nch}} = \frac{\sqrt{ \sum_i (N_{\mathrm{ch}}^{\mathrm{cell},i} - \langle N_{\mathrm{ch}}^{\mathrm{cell}} \rangle)^2 / N_{\mathrm{cell}}^2 }} {\langle N_{\mathrm{ch}}^{\mathrm{cell}} \rangle}, 
\end{equation}
where $N_{\mathrm{ch}}^{\mathrm{cell},i}$ is the multiplicity in cell $i$ and $\langle N_{\mathrm{ch}}^{\mathrm{cell}} \rangle$ is the event-average multiplicity per cell. The additional factor of $1/N_{\mathrm{cell}}$ ensures the value remains bounded between 0 and 1 to maintain an intuitive physical interpretation analogous to other event-shape observables. The acceptance spans the pseudorapidity intervals $-3.7 < \eta < -1.7$ and $2.8 < \eta < 5.1$, corresponding to ALICE V0 detector coverage. This wide, segmented acceptance is essential to capture the global event topology and minimize bias from localized hard processes. We have used $1 - \rho_{\mathrm{nch}}$ as the definition of falttenicity throughout the analysis, where values approaching 1 correspond to events with minimal multiplicity fluctuations between cells, indicating a uniform  isotropic distribution of particles characteristic of soft processes and high MPI activity. Conversely, values approaching 0 signify events with large multiplicity fluctuations, where particle production is highly localized in a few cells, characteristic of jet-like events dominated by hard scatterings. Compared to traditional multiplicity-based classifiers like the V0M estimator, flattenicity demonstrates a comparable sensitivity to the MPI activity but crucially exhibits a significantly reduced bias towards hard processes~\cite{ALICE:2024vaf}, thereby selecting event samples with softer transverse momentum spectra and weaker jet correlations in angular correlation studies. Its robustness has been verified against variations in the granularity of the $\eta$-$\varphi$ segmentation, and its application has been shown to enhance characteristic soft-QCD signals, such as the pronounced enhancement of baryon-to-meson ratios at intermediate transverse momenta~\cite{Ortiz:2022mfv}, establishing it as a powerful and feasible tool for probing the underlying event dynamics in the upcoming LHC runs. In this study, MPI and flattenicity are used as global estimators of event activity and geometry. It is hypothesized that because they are more directly correlated with the initial spatial distribution of partonic interactions and the density of color strings, they will provide a clearer and potentially less biased mapping to the observed long-range correlation strength $V_{2\Delta}$ than the traditionally used mid-rapidity charged-particle multiplicity $N_{\rm ch}$.

\section{Results}
\label{results}

For this analysis, the kinematic selections for the calculation of ultra-long-range azimuthal correlations are as follows. The event multiplicity classification is primarily based on the number of charged particles, $N_{\rm ch}$ within the pseudorapidity range \(|\eta| < 0.8\) and with transverse momentum \(0.2 < p_{\mathrm{T}} < 3\) GeV/\(c\) as done in the ALICE analysis. The final correlation analysis, specifically the extraction of two-particle correlation coefficients \(V_{2\Delta}\), was performed using particle pairs produced in the forward region. The acceptance cuts for the study of correlations is in the range \(5.0 < |\Delta\eta| < 6.0\), combining particles from \(2.6 < \eta < 3.2\) and \(-3.0 < \eta < -2.0\) as done in the ALICE analysis.

\begin{figure}[H]
    \centering
\includegraphics[width=0.95\linewidth]{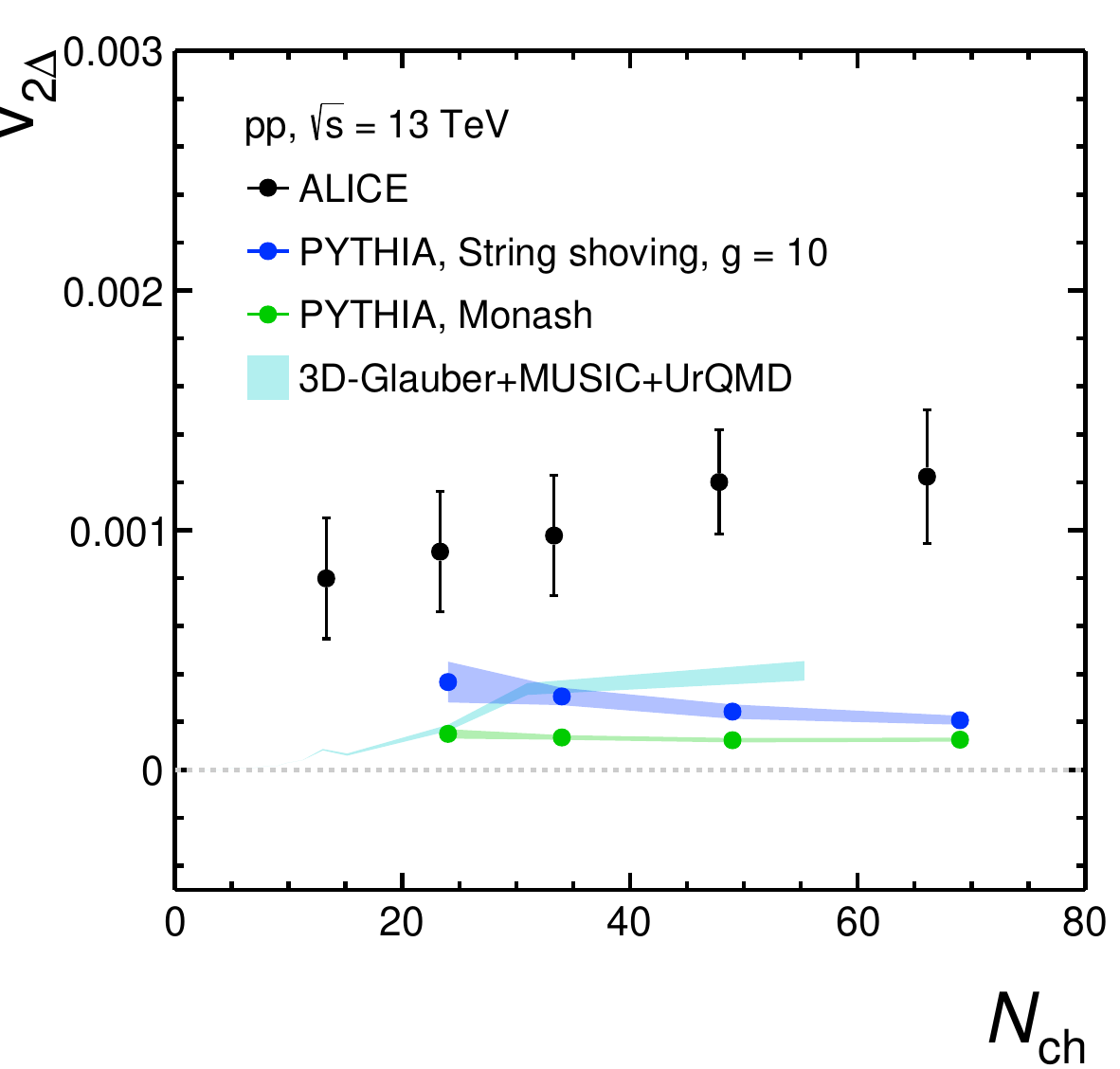}
\caption{The $N_{\rm ch}$ dependence of the second-order two-particle correlation coefficient $V_{2\Delta}$ in pp collisions at $\sqrt{s}$ = 13 TeV. The black markers represent the ALICE data~\cite{ALICE:2025bwp} and the cyan band is for the hydro simulations~\cite{Zhao:2022ugy}. The shaded band for the PYTHIA results are the systematic uncertainties, whereas the bars are for statistical error.}
    \label{fig1}
\end{figure}

Following the template fit procedure mentioned in the methodology section, the $V_{2\Delta}$ parameter is extracted as a function of $N_{\rm ch}$ using PYTHIA8 simulations with string shoving mechanism, as shown in Fig~\ref{fig1}. The multiplicity classes are shown in table~\ref{tab2}. For a baseline comparison, $V_{2\Delta}$ is also extracted by considering PYTHIA8 with Monash tune, which ideally has no flow contribution in it. For Monash, one can observe a flat $V_{2\Delta}$ as a function $N_{\rm ch}$, although the values are minuscule. On the other hand, for the string shoving with the shoving parameter $g$ = 10, one can observe a higher value of $V_{2\Delta}$ across all $N_{\rm ch}$ as compared to Monash results. However, the $V_{2\Delta}$ values decrease as a function of $N_{\rm ch}$ and at higher multiplicity, both the results from Monash tune and string shoving seem to be converging, which means the effect of string shoving vanishes. This decreasing trend in string shoving mechanism has also been observed in Ref.~\cite{Kim:2021elv}. The reason for such trend is supposed to be the large rapidity gap used in the study, as shown in Ref.~\cite{Bierlich:2018lbp}, where without the large rapidity gap the trend from string shoving seems to be flat or increasing.

Statistical uncertainties on the extracted \(V_{2\Delta}\) are computed from the covariance matrix of the template fit. Due to the large event samples, these uncertainties are extremely small ($\sim 1\%$ relative error) and are represented by error bars in the figures, though they often fall within the marker size. Systematic uncertainties are evaluated from three independent sources: the low-multiplicity template definition, the $|\Delta\eta|$ range, and the $\Delta\phi$ fit range. For the low-multiplicity template definition, in addition to the nominal choice ($N_{\mathrm{ch}} = 0$--$10$), two alternative intervals are considered: $N_{\mathrm{ch}} = 0$--$8$ and $N_{\mathrm{ch}} = 0$--$12$. For the $|\Delta\eta|$ range, the nominal selection $5.0 < |\Delta\eta| < 6.0$ is varied to $4.8 < |\Delta\eta| < 5.8$ and $5.2 < |\Delta\eta| < 6.2$. For the $\Delta\phi$ fit range, the nominal range is varied from $-\pi/2$ to $3\pi/2$  to $-\pi/2 + 0.5$ to $3\pi/2 - 0.5$. For each multiplicity bin and each source, we take the maximum and minimum $V_{2\Delta}$ values obtained across the variations. The total systematic uncertainty is then calculated as $\sigma_{\mathrm{sys}} = \frac{V_{2\Delta,\mathrm{max}} - V_{2\Delta,\mathrm{min}}}{\sqrt{12}},$ corresponding to the standard deviation of a uniform distribution over the observed variation, where the index $i$ runs over the three sources. This provides a conservative estimate of the combined systematic effects.


In the figure, the PYTHIA8 results are also compared to ALICE data and to a hydro model which uses initial conditions from Glauber model, incorporating MUSIC hydrodynamic simulations and later UrQMD transport model for the hadrons~\cite{Zhao:2022ugy}. This hydro model gives an increasing $V_{2\Delta}$ trend with increasing $N_{\rm ch}$, however it still largely underestimates the ALICE data. The comparison of ALICE data to models provide support for the emerging understanding of a gradual onset of collectivity as a function of system size and density~\cite{Grosse-Oetringhaus:2024bwr}. The observed dominance of the string shoving mechanism, which is a non-fluid dynamic and initial-state effect, in generating azimuthal correlations in low-multiplicity pp collisions demonstrates that collective-like effects can emerge without a thermalized medium. On the other hand, the hydrodynamic model is incapable of explaining the low multiplicity regime, but has a more prominent effect at the highest multiplicities. This transition between these two mechanisms, from initial-state string interactions in small, sparse systems to final-state collective expansion in larger, denser environments, offers a concrete picture in which the relative contributions of non-fluid and fluid-dynamics evolve smoothly with system size. 

\begin{figure}[h!]
    \centering
\includegraphics[width=0.95\linewidth]{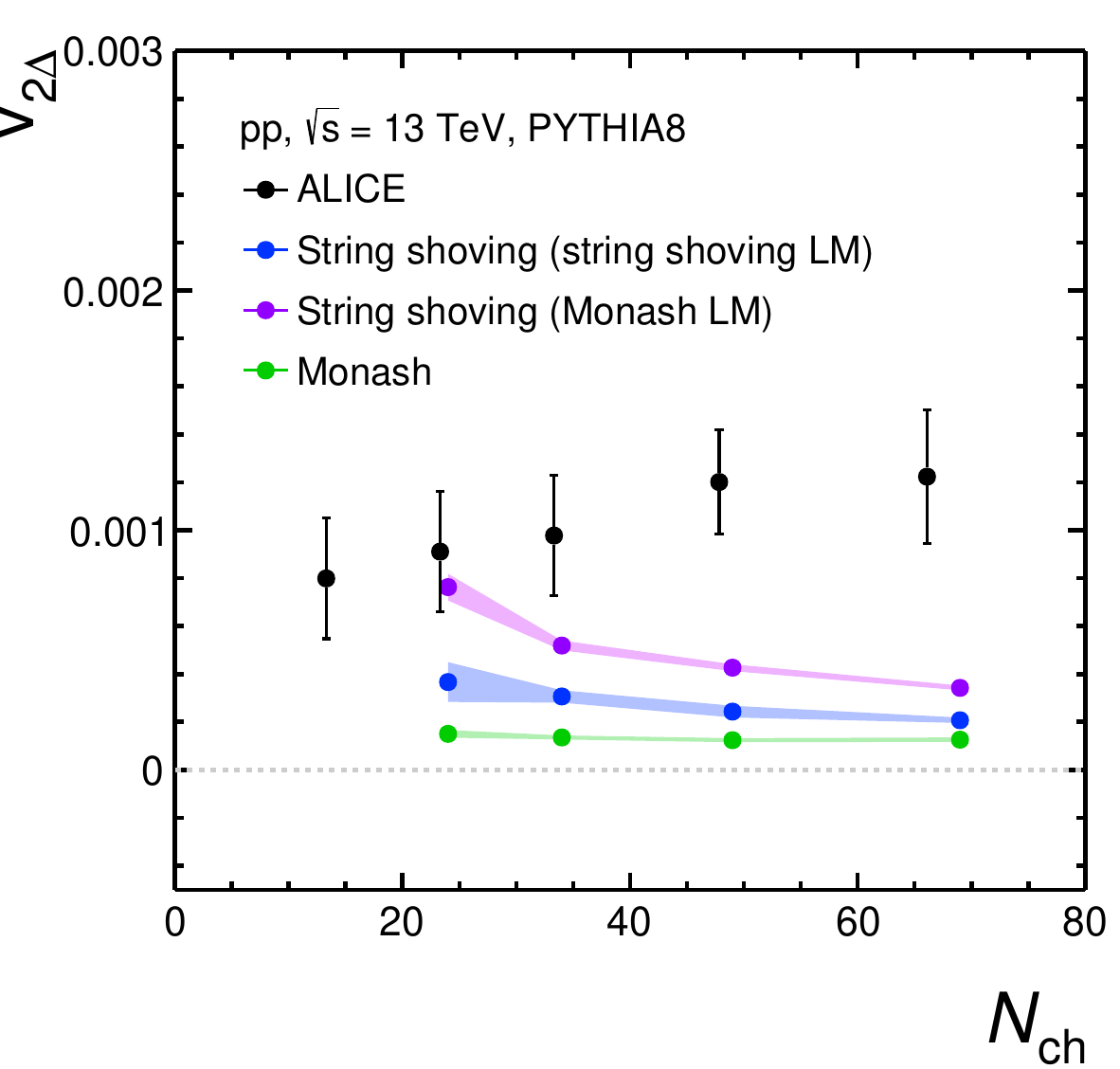}
    \caption{The $N_{\rm ch}$ dependence of the second-order two-particle correlation coefficient $V_{2\Delta}$ in pp collisions at $\sqrt{s}$ = 13 TeV using PYTHIA8 simulations. The black markers represent the ALICE data~\cite{ALICE:2025bwp}. The shaded band for the PYTHIA results are the systematic uncertainties, whereas the bars are for statistical error.}
    \label{fig2}
\end{figure}

\begin{table*}[t]
\centering
\caption{Comparison of $\langle {\rm d}N_{\text{ch}}/{\rm d}\eta\rangle$ for different event classifiers in pp collisions at $\sqrt{s} = 13$ TeV. The flattenicity values correspond to ($1-\rho_{nch}$) as defined in Eq. (4).}
\label{tab2}
\small
\renewcommand{\arraystretch}{1.5}
\setlength{\tabcolsep}{15pt}
\begin{tabular}{|c|c|c|c|c|c|c|}
\hline
{\bf Class} & {\bf $N_{\text{ch}}$} & {\bf $N_{\text{mpi}}$} & {\bf Flattenicity} & {\bf $N_{\text{ch}}$} & {\bf $N_{\text{mpi}}$} & {\bf Flattenicity}  \\
    & {\bf Range} & {\bf Range} & {\bf Range} & {\bf $\langle {\rm d}N_{\text{ch}}/{\rm d}\eta\rangle$} & {\bf $\langle {\rm d}N_{\text{ch}}/{\rm d}\eta\rangle$} & {\bf $\langle {\rm d}N_{\text{ch}}/{\rm d}\eta\rangle$} \\
\hline
I & 0--10 & 0--3 & 0.00--0.78 & 4.10 & 6.05 & 6.82 \\
\hline
II & 10--20 & 3--7 & 0.78--0.80 & 9.55 & 10.73 & 11.70 \\
\hline
III & 20--30 & 7--11 & 0.80--0.82 & 15.52 & 15.52 & 16.60 \\
\hline
IV & 30--40 & 11--14 & 0.82--0.83 & 21.54 & 19.21 & 20.04 \\
\hline
V & 40--60 & 14--18 & 0.83--0.84 & 29.19 & 22.24 & 22.70 \\
\hline
VI & 60--80 & 18--19 & 0.84--1.00 & 40.99 & 25.40 & 27.86 \\
\hline
\end{tabular}
\end{table*}

In Fig.~\ref{fig2}, the estimated $V_{2\Delta}$ from PYTHIA8 simulations in different scenarios is shown. First, the low multiplicity (LM) event activity class is chosen as the baseline template and the rest of the higher multiplicity classes are fitted, 
everything estimated from the string shoving mechanism. For the second case, the low multiplicity template from the Monash tune is chosen and fitted to the higher activity classes from string shoving mechanism. Finally, the Monash tune low multiplicity template is used to fit the higher multiplicity classes from the same Monash results. The motivation for this particular analysis is very simple. The Monash tune is particularly valuable as a non-flow reference in studies of collective-like phenomena because it incorporates conventional QCD processes, including multiparton interactions, parton showers, and string fragmentation, without mechanisms that generate collective anisotropic flow, such as hydrodynamic expansion or string repulsion. Thus, the low multiplicity template from Monash can act as a pure non-flow template which can be subtracted from the rest of the higher $N_{\rm ch}$ classes. Indeed it is observed that when the low-multiplicity event class from Monash is used as the template, a higher magnitude of $V_{2\Delta}$ is extracted, as shown by the violet markers. For this case, the estimation agrees well with the experimental data within uncertainties for the lowest multiplicity class, although the trend is still decreasing, and it still underestimates the data at higher multiplicities. This result points to the fact that, the low-multiplicity template from string shoving already contained some flow-like effects which reduces the $V_{2\Delta}$ magnitude for the higher event classes after template fit. Such observations have also been made in Ref.~\cite{Kim:2021elv} where the authors show that when the low-multiplicity template is scaled and subtracted from the higher multiplicity signal, it results in an over-subtraction, artificially suppressing the extracted $V_{2\Delta}$ values. In the string-shoving mechanism, each overlapping string pair provides a small anisotropic transverse push. As the event becomes more active (with an increasing number of strings), two different effects might reduce the estimated $v_2$. Firstly, the average cancellation of many randomly oriented push, and 
secondly the dilution of the anisotropy when the same total anisotropic momentum is shared among a larger number of particles. Furthermore, the template subtraction procedure can amplify the apparent decrease of $v_2$ by over-subtracting any residual long-range component present in the low-multiplicity reference. These combined effects naturally lead to a decreasing trend of $v_2$ with $N_{\text{ch}}$, even when an absolute long-range ridge structure remains visible. These results somewhat confirm that string shoving mechanism alone cannot generate enough flow-like effects in the system at very long range. Moreover, while these studies support the robustness of the extraction procedure, a fully controlled injection/recovery closure test within a dynamically generated Monte Carlo framework remains nontrivial and is left for future investigation.

It should also be noted that the use of midrapidity $N_{\text{ch}}$ as an event classifier introduces a bias, since it preferentially selects events with enhanced jet or non-flow activity, thereby diluting the genuine collective component. For further study, two other event estimators are selected along with $N_{\text{ch}}$; the number of multiparton interactions $N_{\rm mpi}$, and flattenicity. This choice is also physically motivated, as the string shoving mechanism is fundamentally governed by the density and spatial configuration of color strings in the initial state. For this, the $N_{\rm mpi}$ serves as a more direct proxy for the number of primary color strings created in the collision. As MPI is not accessible via the experiments, one can further use flattenicity which characterizes the global event topology. Therefore, by choosing $N_{\rm mpi}$ and flattenicity, one is using observables that are more directly correlated with string density that the shoving mechanism is postulated to depend on, in addition to reduce the $N_{\text{ch}}$ bias. This allows for a cleaner and more physically meaningful test of the model.

Table~\ref{tab2} provides the event class definitions for all the three event classifiers and their corresponding average charged-particle densities, $\langle {\rm d}N_{\text{ch}}/{\rm d}\eta\rangle$. The event classes for $N_{\text{ch}}$, $N_{\rm mpi}$, and flattenicity are constructed to contain an identical fraction of the total event sample. This ensures a direct comparison between classifiers, where differences in the observed $V_{2\Delta}$ can be attributed to the distinct physical properties each estimator captures, rather than trivial differences in event statistics. The resulting $\langle {\rm d}N_{\text{ch}}/{\rm d}\eta\rangle$ values reveal the different event topologies selected by each classifier. For the same event class index (i.e., the same fraction of events), the $N_{\text{ch}}$ classifier yields the lowest average multiplicity in the lowest activity class and the highest in the highest activity class. This reflects its sensitivity to local particle yield, which can be high due to either many multi-parton interactions or a single hard jet. In contrast, $N_{\rm mpi}$ and flattenicity, which are more global measures, produce significantly different average charged particle densities. Notably, in the highest activity class (class VI), the $N_{\text{ch}}$ classifier selects events with $\langle {\rm d}N_{\text{ch}}/{\rm d}\eta\rangle \approx 41$, while $N_{\rm mpi}$ and flattenicity select events with multiplicities of approximately 25 and 28, respectively. This divergence underscores that a high $N_{\text{ch}}$ value selects a different, often more jet-biased, population of events compared to high-$N_{\rm mpi}$ or high-flattenicity selections, which are more representative of a dense, isotropic underlying event.

\begin{figure*}
    \centering
\includegraphics[width=0.95\linewidth]{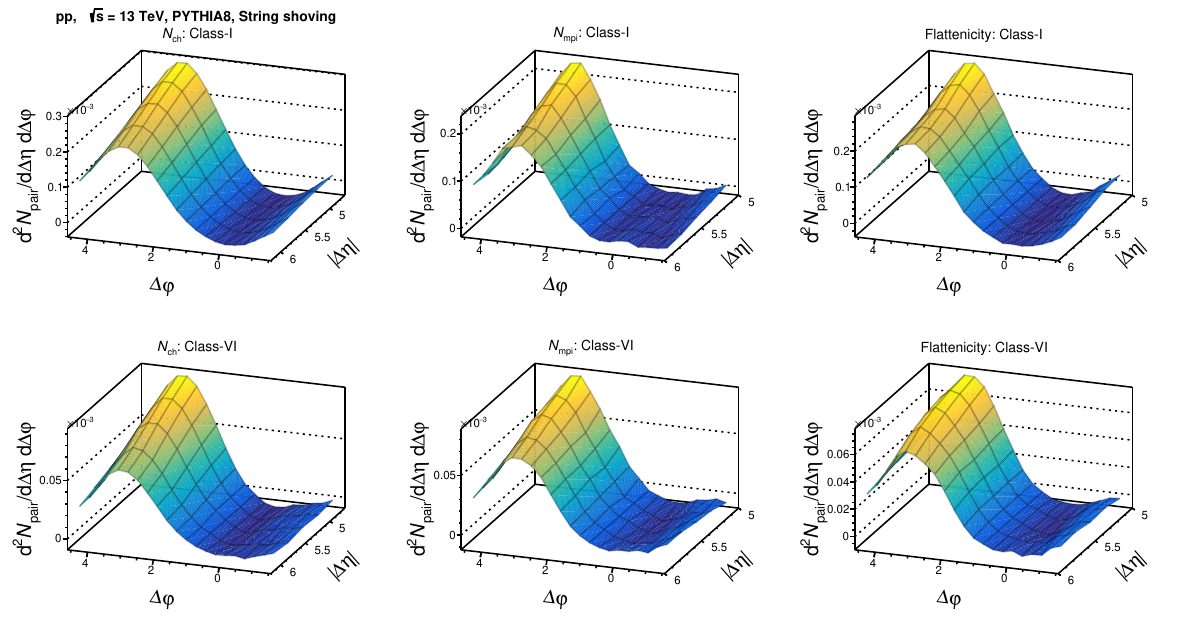}
    \caption{The ultra-long range correlation yield estimated as a function of $\Delta\eta$ and $\Delta\varphi$ in pp collisions at $\sqrt{s}$ = 13 TeV with PYTHIA for different estimators for small (class-I) and large event activity (class-VI). The flattenicity definition corresponds to ($1-\rho_{nch}$) as defined in Eq. (4).}
    \label{fig3}
\end{figure*}

\begin{figure*}
    \centering
\includegraphics[width=0.95\linewidth]{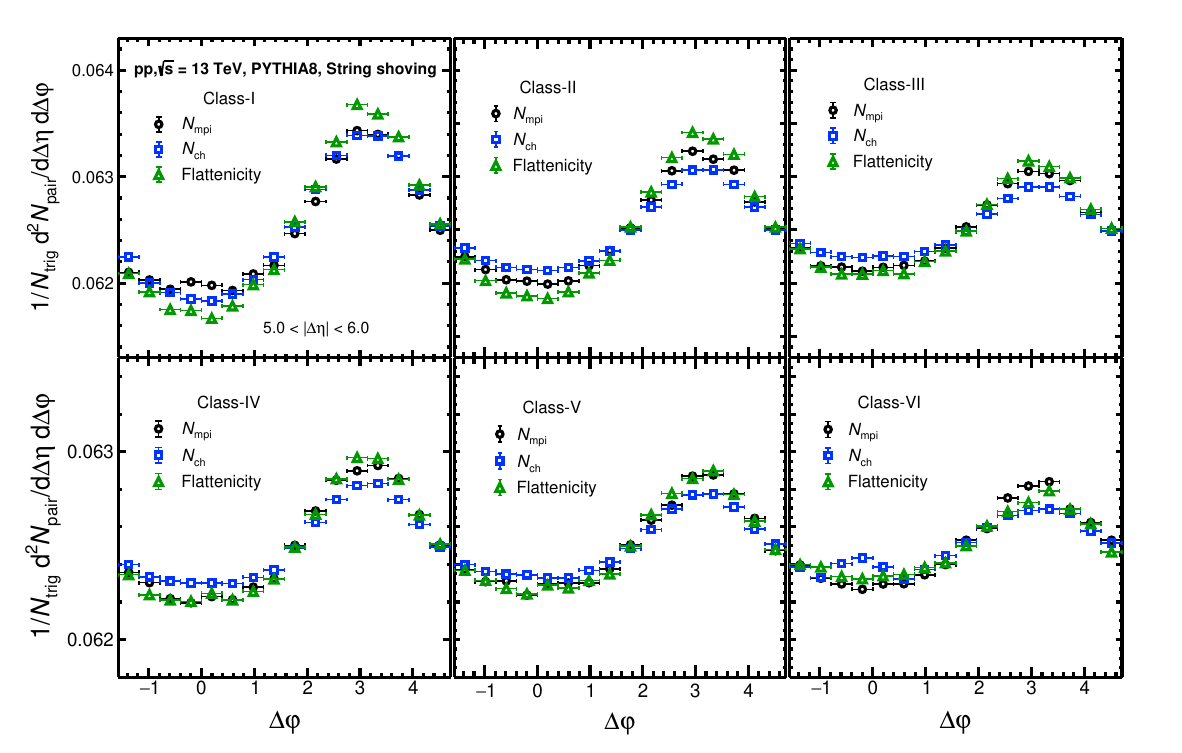}
    \caption{The ultra-long range correlation yield estimated as a function of $\Delta\varphi$ in pp collisions at $\sqrt{s}$ = 13 TeV with PYTHIA for different estimators. The flattenicity definition corresponds to ($1-\rho_{nch}$) as defined in Eq. (4).}
    \label{fig4}
\end{figure*}

Figure~\ref{fig3} shows the two-particle correlation function $C(\Delta \eta,\Delta\varphi)$ for low- and high-event activity classes in all three different estimators, namely $N_{\rm ch}$, $N_{\rm mpi}$ and flattenicity. To emphasize the azimuthal correlation relative to the baseline, each distribution is normalized by its integral and then shifted such that the minimum yield is set to zero (ZYAM procedure). This convention allows for visual comparison of the near-side ridge structure across different event classes and estimators. Although the ridge structure is not very prominent as such large $\eta$, one can still see small structures in the near side region for the low-$N_{\rm mpi}$ and high-$N_{\rm ch}$ event classes. To properly visualize this, Fig.~\ref{fig4} shows the $\Delta\varphi$ distributions for different event classes using all three estimators. As one can observe, for the lowest-$N_{\rm ch}$ class, which is taken as the low multiplicity template, there is no structure in the near side region, whereas the lowest-$N_{\rm mpi}$ class, which corresponds to the same fraction of events as the lowest-$N_{\rm ch}$ class, shows a substantial structure in the near side region.
Similar to $N_{\rm ch}$, flattenicity also does not show any structure in the lowest-event class. This stark contrast in the near-side correlation structures between different event classes even at the same fraction of total events, stems from their inherent sensitivity to different underlying event topologies.

Contrary to the naive expectation that collective flow requires high-event activity, the string shoving mechanism predicts that the strongest long-range azimuthal correlations can arise in events with a simple, elliptical geometry of a few overlapping strings. This observation aligns with the model expectation, where a discernible ultra-long-range ridge structure is seen in low-$N_{\rm mpi}$ events. In such sparse environments, the limited number of color strings whose positions are influenced by the geometry of the beam remnants can create a configuration with sufficient initial anisotropy for the string shoving mechanism to operate efficiently. The traditional $N_{\rm ch}$ estimator fails to isolate these geometrically preferential events, as it mixes them with relatively high-$N_{\rm mpi}$ events (which are more isotropic) and pencil-like jet events (which contribute only short-range correlations). This is consistent with the string shoving mechanism operating most effectively in a specific, geometrically anisotropic regime that $N_{\text{ch}}$ largely fails to select. This result is consistent with recent proposals that string repulsion in sparse systems can generate negative elliptic flow, an enhancement along the major axis of the initial ellipse, contrasting with the positive flow from hydrodynamics~\cite{Bierlich:2024lmb}. On the other hand, the lowest flattenicity event class is characterized by large relative multiplicity fluctuations across the $\eta-\varphi$ acceptance. Such events lack the extended transverse spatial distribution and the global geometric anisotropy required for the string shoving mechanism to generate a significant collective push, as the repulsive interaction depends on the overlap of multiple color strings in the transverse plane. This distinction highlights that flattenicity is not merely an alternative to multiplicity estimator, but a genuine event-shape observable that cleanly separates jet-dominated topologies from the isotropic, MPI-rich environments where collective behavior emerges. The absence of near-side structure in low-flattenicity events suggests that the ridge observed in low-$N_{\rm mpi}$ events is not arising from jet fragmentation, rather is a signature of the string shoving mechanism, which requires a specific, anisotropic spatial configuration of initial-state color fields, including those from beam remnants, to generate a repulsive transverse push.

\begin{figure*}
    \centering
\includegraphics[width=0.95\linewidth]{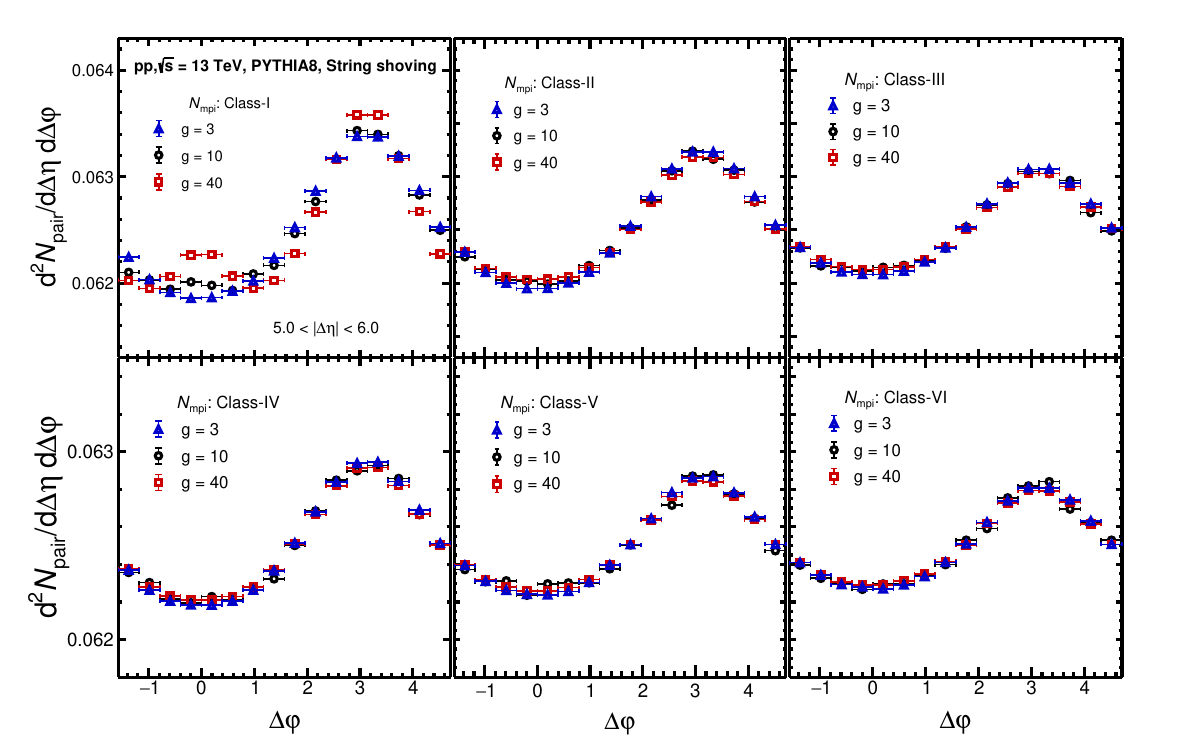}
    \caption{The ultra-long range correlation yield estimated as a function of $\Delta\varphi$ in pp collisions at $\sqrt{s}$ = 13 TeV with PYTHIA for different MPI event activity using different values of string shoving parameter.}
    \label{fig5}
\end{figure*}

Moreover, as one goes to the highest $N_{\rm ch}$ class, one observes a substantial structure in the near side region whereas the same is absent for highest $N_{\rm mpi}$ and flattenicity classes. For higher event activity, flattenicity and $N_{\rm mpi}$ seem to be showing similar behavior in the distribution. The substantial near-side structure observed in the highest $N_{\rm ch}$ event class, which is absent in the highest $N_{\rm mpi}$ and flattenicity classes, suggests that $N_{\rm ch}$ introduces a significant selection bias. While the ultra-long-range nature of the correlation rules out a simple jet fragmentation origin, the high-$N_{\rm ch}$ selection likely favors events with a particular initial-state geometry that is already predesigned to yield a large number of particles in the mid-rapidity acceptance. This geometric bias, potentially coupled with correlations triggered by semi-hard processes, artificially enhances the observed correlation signal. In contrast, the $N_{\rm mpi}$ and flattenicity classifiers, which are more directly tied to the global event topology and are less sensitive to such momentum-based biases, reveal the underlying trend predicted by the string shoving model, that the efficiency of the mechanism saturates or decreases in very dense, symmetric environments.

Finally, the dependence of the azimuthal correlation structure on both the number of MPI and the string shoving strength parameter \(g\) is shown in Fig.~\ref{fig5}. In the lowest MPI class, a clear near-side structure is seen which gets enhanced with the increase in shoving parameter (\(g=40\)) compared to the moderate case (\(g=10\)), whereas the structure is not present at a very small shoving parameter ($g=3$). This demonstrates that the string shoving mechanism is most effective in sparse environments with few overlapping strings, where initial geometric anisotropies are large and not washed out by cancellations. However, this distinctive structure diminishes and becomes indistinguishable between the two \(g\) values as \(N_{\text{mpi}}\) increases across subsequent classes. The correlation does not show any significant azimuthal modulation in the highest activity class. This provides evidence for one key conclusion of our study that the string shoving mechanism possesses a built-in saturation where its efficiency peaks in geometrically preferential, low-$N_{\rm mpi}$ events and becomes ineffective in dense, isotropic environments. The observed insensitivity to the shoving strength \(g\) at high multiplicities underscores that repulsion between static color strings alone cannot generate the persistent collective behavior observed in experimental data.

It is important to emphasize that $N_{\mathrm{mpi}}$ is an initial-state property determined at the parton level before any final-state evolution, including parton showers, hadronization, and string shoving. Classifying events by $N_{\mathrm{mpi}}$ therefore introduces no bias from final-state effects and provides a clean proxy for the density of color strings in the initial collision. In addition, unlike hydrodynamic models that describe the evolution of a thermalized medium, PYTHIA8 with the string-shoving mechanism contains no collective flow from a QGP. All azimuthal correlations in this framework arise from collective-like effects, specifically the repulsive transverse pressure between overlapping color strings. Thus, the $V_{2\Delta}$ extracted from these simulations is a collective-like correlation arising from string shoving. This is a crucial distinction when interpreting the comparison with ALICE data and hydrodynamic calculations.

\section{Summary}
\label{summary}

This work presents a systematic investigation of the potential collective phenomena in small systems through ultra-long-range correlations. The PYTHIA8 event generator with string shoving mechanism is used and its ability to describe the azimuthal correlations in pp collisions at $\sqrt{s} = 13$ TeV is tested by extracting the correlation strength $V_{2\Delta}$ in the ultra-long-range region ($|\Delta\eta| > 5.0$) using a template fit method. The model is used as a proxy to understand the selection biases on the extraction of $V_{2\Delta}$. Although we know that the nonzero $V_{2\Delta}$ in the model is a collective-like effect produced by the string shoving, and not from hydrodynamics. The novel contributions of this work are: (i) the first systematic comparison of charged-particle multiplicity ($N_{\mathrm{ch}}$), number of multiparton interactions ($N_{\mathrm{mpi}}$), and flattenicity as event classifiers within the string-shoving framework for ultra-long-range correlations; (ii) the identification of a built-in saturation mechanism in the string-shoving model, where the strongest signal appears in low-$N_{\mathrm{mpi}}$ (dijet-like) events and diminishes in dense, isotropic environments; and (iii) the demonstration that flattenicity selects cleaner correlation structures with reduced bias from hard processes compared to traditional $N_{\mathrm{ch}}$-based classification. In addition, the principal findings of this analysis are as follows.

 \begin{enumerate}


\item Contrary to the intuitive link between collectivity and high activity, the string-shoving mechanism generates its strongest ultra-long-range correlations in events with a low number of multiparton interactions ($N_{\rm mpi}$ $\approx$ 1–3). In these dijet-dominated events, the anisotropic geometry of a few overlapping strings and beam remnants allows for an efficient collective push.

\item The traditional event classifier $N_{\text{ch}}$ introduces a significant bias. Using $N_{\text{ch}}$ as an event-activity estimator conflates events with genuine underlying-event activity and those with high particle yield from jets. This bias dilutes the collective-like signal and artificially exacerbates the decreasing $V_{2\Delta}$ trend, worsening the apparent failure of the model.

\item Global event-shape estimators provide a clearer view of the model behavior. When using direct proxies for string density like the number of MPIs or the novel event-shape observable like flattenicity, the analysis shows that the string-shoving signal saturates and diminishes in dense, isotropic environments. This built-in saturation is a key feature of the mechanism.

\item The template fit method is further complicated by inherent flow-like effects in low-activity events. It is demonstrated that the low-multiplicity templates used for non-flow subtraction already contain significant flow-like correlations from string shoving. This leads to over-subtraction in higher activity classes, further suppressing the extracted $V_{2\Delta}$.

\item The analysis results support a picture of a gradual onset of collectivity as a function of system size and density. In this picture, initial-state effects like string shoving dominate in low-multiplicity pp collisions, while final-state collective expansion (hydrodynamics) becomes increasingly relevant in high-multiplicity pp collisions.

 \end{enumerate}

In conclusion, this work strongly advocates for moving beyond $N_{\text{ch}}$-based analysis in favor of global, topology-sensitive estimators like Flattenicity to achieve a less biased and more physically meaningful interpretation of collective phenomena in small collision systems.

\section*{Acknowledgment}
Authors acknowledge the technical support from Luciano D\'iaz Gonz\'alez and Jes\'us Eduardo Murrieta Le\'on. This work has been supported by DGAPA-UNAM PAPIIT No. IG100524 and PAPIME No. PE100124. G.B. acknowledges the support provided by the Hungarian National Research, Development and Innovation Office under the grants $\rm OTKA~ PD\_22~ 143226, NKKP~ADVANCED\_25~ 153456$, $\rm 2021-4.1.2-NEMZ\_KI-2024-00035$.

\end{document}